\documentstyle [11pt,psfig]{article}
\begin{document}
\begin{center}{
{\bf  On the Variational method for LPM Suppression of Photon Emission from Quark-Gluon Plasma }
\\
{S. V. S. Sastry}\\
{ Nuclear Physics Division, Bhabha Atomic Research Centre,
Trombay, Mumbai 400 085, India}
\vspace {-.4cm}
}\end{center}

\par
\vspace {1.5cm}
\centerline{Abstract}
\par

\noindent
The  photon  emission  rates  from the quark gluon plasma have been studied
considering  LPM  suppression  effects.  The  integral  equation  for   the
transverse  vector  function  (${\bf  \tilde{f}(\tilde{p}_\perp  )}$)  that
consists  of  multiple   scattering   effects   has   been   solved   using
self-consistent  iterations method. Empirical fits to the peak positions of
the  distributions  from  iteration   method   have   been   obtained   for
bremsstrahlung  and $\bf aws$ processes. The variational approach for ${\bf
\tilde{f}(\tilde{p}_\perp )}$ calculation has been simplified  considerably
making  some  assumptions.  Using this method, the photon emission rates at
finite baryon density have been estimated. The LPM suppression factors  for
bremsstrahlung  and $\bf aws$ processes have been obtained as a function of
photon energy and baryon density. The effect of  baryon  density  has  been
shown to be rather weak and the suppression factors are similar to the zero
density  case. The suppression factors for $\bf aws$ processes can be taken
at zero density, whereas the bremsstrahlung suppression  can  be  taken  at
zero density multiplied by a density dependent factor.

\par

\vspace {2.cm}

\noindent
Electromagnetic  processes  such  as  photons  and dileptons production are
considered to be important signals of formation of the quark  gluon  plasma
(QGP)  in  the  relativistic heavy ion collisions. The experimental data by
WA98 collaboration \cite{wa98} of direct photons from Pb+Pb  collisions  at
CERN  SPS  is  of paramount importance which boosted interest in the static
emission  rates  and  the   convoluted   photon   yield   calculations.   A
comprehensive  review  on  the  theoretical and experimental studies of the
photon emission has been given in \cite{peitz}. The photon  emission  rates
from hot hadron gas, the QGP phase and also the prompt photon emission have
been  investigated  in  great  detail (see references in \cite{peitz}). The
processes contributing to photon emission from QGP phase at (HTL) effective
one loop level are the quark-antiquark annihilation into  a  photon  and  a
gluon and the absorption of a gluon by a quark (anti quark) emitting photon
(see  Fig.1  of  \cite{peitz}).  The  processes of bremsstrahlung and ${\bf
aws}$ that arise at effective two loop level contribute at the  same  order
as  one  loop  processes  \cite{auren1}.  These  processes  (see  Fig. 5 in
\cite{peitz}) contribute at the leading order $O(\alpha\alpha_s)$ owing  to
the  collinear  singularity  that  is  regularized by the effective thermal
masses. The higher order multiple scatterings however can not be ignored as
these may also contribute at the  same  order  as  the  one  and  two  loop
processes   \cite{auren2,auren3,auren4,arnold1,arnold2,arnold3}.   Further,
multiple soft scatterings of the fermion during photon emission reduce  the
photon  coherence lengths, known as Landau-Pomeranchuk-Migdal (LPM) effect.
The  photon  emission  rates  are  suppressed  owing  to  the  LPM  effects
\cite{auren3,auren4,arnold1,arnold2}.  Using  the hydrodynamical models for
the expansion of the plasma and convoluting the expansion with  the  photon
emission  rates, one obtains the photon transverse momentum spectrum. These
space time evolution of the plasma through QGP, mixed and hadron phases  is
complicated  if  the  plasma  is  formed  at  finite  baryon density. These
physical conditions of the plasma affect both  the  basic  photon  emission
rate  and  the  evolution  of the plasma and thus the space time integrated
photon spectrum. In view of this, the formalism of Aurenche {\it et.  al.,}
of  photon emission rates at two loop level has been recently extended to a
chemically unsaturated plasma at finite baryon density \cite{dsmkr1}. These
emission rates (meaning same as production rates  in  this  work)  and  the
space  time  (3+1  dimensional)  expansion dynamics of the plasma at finite
baryon density have been recently reported \cite{dsmkr2,sdms1}. The  photon
spectra  have  been  shown  for  various  cases  of plasma at finite baryon
density for SPS and RHIC energies and the baryon free case in  \cite{sdms1}
included the LPM effects.

\noindent

\par

\noindent
The photon production rates from bremsstrahlung and the $\bf aws$ processes
have  been  estimated  by  Aurenche  {\it  et.  al.,}  \cite{auren1}.  In a
remarkable simplification, these rates are expressed in terms of simple one
dimensional momentum integrals and the dimensionless  quantities  $J_T,J_L$
\cite{auren1}.  These  values  are  not  very  sensitive  to baryo chemical
potential and they weakly depend  on  the  thermal  masses  for  gluon  and
fermions  through $m_g^2/m_\infty^2$ \cite{dsmkr1}. The differential photon
emission rate per unit volume of energy $k_0$ is given by (denoted by${\cal
R}^0$ for simplicity of later use),

\begin{equation}
{{\cal R}^0}_{b,a} =
\frac{40\alpha\alpha_s}{9\pi^4}n_B(k_0) \frac{T}{k^2}(J_T-J_L)~I_{b,a}(k)
\end{equation}

\begin{equation}
I_{b}(k) = \int_0^\infty dp\left(p^2+(p+k)^2\right) \left[(n_f(p)-n_f(p+k)) + (\bar{n}_{f}-\bar{n}_{f}(p+k)\right]
\end{equation}

\begin{equation}
I_{a}(k) = \int_0^\infty dp\left(p^2+(k-p)^2\right) \left[1-\bar{n}_{f}(p)-\bar{n}_{f}(k-p)\right]
\end{equation}

\noindent
In  above,  the subscripts ~$(b,a)$~ denote the bremsstrahlung and the $\bf
aws$ processes  and  we  consider  a  two  flavor  three  color  case  with
$\alpha_s=0.2$. These rates in Eq. 1 do not include LPM suppression effect.
The effects of finite baryon density on the photon emission rates have been
discussed in \cite{dsmkr2} and these results are summerised as follows. The
bremsstrahlung  radiation  is  affected  whereas the ${\bf aws}$ process is
insensitive to the baryon density presence.  The  bremsstrahlung  radiation
from  quark  is enhanced and from anti quark is suppressed at finite baryon
density. As mentioned earlier, it has been shown that multiple  scatterings
during the photon emission for the bremsstrahlung and ${\bf aws}$ processes
cannot  be ignored. The exchange of a soft gluon of finite thermal mass and
the collinear singularity (arising when photon is emitted parallel to quark
momentum) regularized by the thermal masses  are  essentially  the  factors
that  make  the  two loop processes contribute at the same order as the one
loop processes. It has been realized \cite{auren3,arnold1} that for similar
reasons, multiple gluon exchange  processes  in  terms  of  certain  ladder
diagrams  also  contribute  to the same order. The diagrammatic analysis of
all the processes contributing to the same order  has  been  performed  and
these  contributions  have  been  summed  in  \cite{arnold1}. This leads to
suppression of photon emission  rates  as  shown  by  the  LPM  suppression
factors  in  Fig. 7. of \cite{arnold2}. One can also obtain the suppression
of the total emission rate by using the empirical expressions in  Eq.(1.10)
of  \cite{arnold2}  and  the  two  loop  rates  of  Eq. 1 above without LPM
effects. It has been shown that the photon radiation from bremsstrahlung is
strongly  suppressed  to  $\sim  ~0.2$  at  very  low  $k/T$  values.   The
suppression  becomes  weaker at higher photon energies and approaches $\sim
0.86$ for $k/T > 10$. In contrast, the $\bf aws$ is unaffected at low $k/T$
values, but falls almost linearly up to $k/T \sim 10$.  The  bremsstrahlung
from  anti  quark  is  same as that of quark for a baryon free case. In the
presence of finite baryon density, the bremsstrahlung from  quark  will  be
different  from  the  anti  quark.  Therefore,  in  order  to  compare with
experimental  photon  spectra,  it  is  necessary  to  understand  how  the
suppression  factors  change  with  baryon  density. The chemical potential
dependence of the emission rates are contained in the population  functions
of  the quarks (anti quarks). Further, for comparing with experimental data
it is necessary to determine these suppression factors  (beyond  $k/T  \sim
10$)  up to $k/T\sim 20$ {\it i.e.,} $k \sim 5$ GeV for T=0.25GeV case. The
differential photon emission rate (denoted by $\cal R$) that  includes  the
LPM effects is given by (for details see \cite{arnold2}),

\begin{equation}
{\cal R}_{b,a}= \frac{80\pi T^3\alpha\alpha_s}{(2\pi)^3 9\kappa}\int_{-\infty}^{\infty}dp_\parallel
\left[\frac{p_\parallel^2+(p_\parallel+k)^2)}{p_\parallel^2(p_\parallel+k)^2)}\right]n_f(k+p_\parallel)(1-n_f(p_\parallel))
~ 2{\bf \tilde{p}_\perp \cdot\Re\tilde{f}(\tilde{p}_\perp)}
\end{equation}

\noindent
In  the  above  $\kappa=m_\infty  ^2/m_D^2  ~~(=\frac{1}{4} ~~~\mbox{for}~~
\mu=0~~~ \mbox{case})$ and the subscripts $(b,a)$  are  for  bremsstrahlung
and  ${\bf  aws}$  as  in  Eq.  1,  with  different  kinematic  domains and
appropriate distribution functions.  The  value  of  $\kappa  $  is  baryon
density  dependent  and is determined by evaluating these thermal masses as
listed in the Table as a  function  of  $\mu_q$.  In  the  above  equation,
$\Re{\bf  \tilde{f}(\tilde{p}_\perp  )}$  is  the real part of a transverse
vector (amplitude) function which  consists  of  the  LPM  effects  due  to
multiple  scatterings.  This  can  be  taken  as  transverse  vector  ${\bf
(\tilde{p}_\perp  )}$  times  a  scalar  function  of  transverse  momentum
${\tilde{p}_\perp  }$.  The  sign  ~$\tilde{}$~  denotes  the dimensionless
quantities in units of Debye mass $m_D$ as defined in  \cite{arnold2}.  The
function   ${\bf  \tilde{p}_\perp\cdot\Re\tilde{f}(\tilde{p}_\perp  )}$  is
determined by the collision kernels ($\tilde{C}({\bf \tilde{q}_\perp})$) in
terms of the following integral equation.

\begin{equation}
2{\bf \tilde{p}_\perp}=i\delta \tilde{E}({\bf \tilde{p}_\perp},p_\parallel,k)
{\bf \tilde{f}}({\bf \tilde{p}_\perp},p_\parallel,k) +
\int\frac{d^2{\bf \tilde{q}_\perp}}{(2\pi)^2}
\left[{\bf \tilde{f}}({\bf \tilde{p}_\perp},p_\parallel,k)
-{\bf \tilde{f}}({\bf \tilde{p}_\perp+\tilde{q}_\perp},p_\parallel,k)\right]
\tilde{C}({\bf \tilde{q}_\perp})
\end{equation}

\begin{equation}
\tilde{C}({\bf \tilde{q}_\perp})=\kappa\int d\tilde{q}_\parallel d\tilde{q}^0
\delta(\tilde{q}^0-\tilde{q}_\parallel) \frac{1}{\tilde{q}}\left[
\frac{2}{|\tilde{q}^2-\tilde{\Pi}_L(\tilde{q}^0,\tilde{q})|^2}+
\frac{\left(1-(\tilde{q}^0/\tilde{q})^2\right)^2}
{|(\tilde{q}^0)^2-\tilde{q}^2-\tilde{\Pi}_T(\tilde{q}^0,\tilde{q})|^2}\right]
\end{equation}
\begin{equation}
\delta \tilde{E}({\bf \tilde{p}_\perp},p_\parallel,k)=
\frac{kT}{2p_\parallel(k+p_\parallel)}\left[\tilde{p}_\perp^2+\kappa \right]
\end{equation}

\noindent
In  above,  the  $\tilde{\Pi}_L,\tilde{\Pi}_T$  are  the  longitudinal  and
transverse parts of (dimensionless) thermal  self  energies  of  the  gauge
fields  and  $\delta \tilde{E}({\bf \tilde{p}_\perp},p_\parallel,k)$ is the
energy difference between the relevant states  of  the  system  before  and
after  the  photon  emission \cite{arnold2}. Based on the sum rules for the
thermal gluon spectral functions, Aurenche, Gelis and Zaraket  obtained  an
analytical   form   for  the  collision  kernel  as  given  by  Eq.  44  of
\cite{auren4}.  This  enormously  simplifies  the  photon   emission   rate
calculations, as it circumvents the need for evaluating the integral in Eq.
6 to obtain the collision kernel. In the present work, we adopt this simple
analytical form for the collision kernel, with appropriate factors properly
taken  care  of.  We  have computed the integral in Eq. 6 for the collision
kernel and compared with the analytical form  given  in  \cite{auren4}.  We
found these to be agreeing, except near ${\bf \tilde{q}_\perp}=0$ where the
collision  kernel  of  Eq.  6  falls  to  zero and can be fitted to another
analytical form for use in calculations. This result has been surprising as
the analytical form in \cite{auren4} is based on very  general  assumptions
satisfied  for  gluon  thermal  (HTL)  self  energies and sum rules for its
spectral functions. The analytical  form  of  \cite{auren4}  for  the  full
integration  region can also be used in the calculations. The divergence of
kernel in \cite{auren4} at ${\bf \tilde{q}_\perp}$=0 is compensated by the
 vector function difference and the ${d^2{\bf \tilde{q}_\perp}}$ together.
Further, use of rationalized momenta in terms  of  debye  mass  cancel  the
$3m_g^2$  terms  in  the  analytical form given by Aurenche {\it et. al.,}.
Therefore, the $\tilde{C}({\bf \tilde{q}_\perp})$ function just  scales  as
$\kappa$  times  the  analytical  forms  which  are independent of chemical
potentials.

\noindent
The  function ${\bf \tilde{p}_\perp\cdot\Re\tilde{f}(\tilde{p}_\perp )}$ as
a function of $|{\bf \tilde{p}_\perp}|$ has to be solved self  consistently
for  each  set of \{$p_\parallel,k,\kappa,T$\} values. In the present work,
we have solved the  equation  by  iterations  at  a  fixed  temperature  of
T=0.25GeV  and  obtained  $2{\bf\tilde{p}_\perp}\cdot\Re{\bf\tilde{f}}({\bf
\tilde{p}_\perp})$ distribution. For small $p_\parallel $  and  $k$  values
the  iterations  converge  very  fast. For small $k$ and large $p_\parallel
(>2.5)$ the convergence is slow. The peak positions of these  distributions
have  been  obtained  for  each set of $\{p_\parallel ,k,\kappa \}$. It has
been noticed that the peak positions of iteration  method  ($A_i$)  can  be
very well approximated by,

\begin{equation}
A_i(p_\parallel ,k) = \left|\frac{1}{p_\parallel}-\frac{1}{p_\parallel+k}\right|^\beta
\end{equation}

\noindent
The  value  of  $\beta$ parameter is, $\beta=-0.32 $ for bremsstrahlung and
$\bf aws$ away from the $p_\parallel$ limits. (The value  of  $\beta=-0.16$
for $\bf aws$ near the $p_\parallel$ limits has been a good approximation.)
Further, a lowest cut-off value of $A_i^{min}=0.32$  was  taken,  which  is
necessary  near  $p_\parallel$  limits or very small $k$ values, especially
for $\bf aws$ case. The ${\bf \tilde{f}}$ distributions have  been  studied
for various values of quark chemical potentials ranging from 0-2GeV. It has
been  found  that the peak positions are not very sensitive to the chemical
potentials. The dependence  of  the  chemical  potential  is  contained  in
$\kappa$  factor of the integral equation which weakly depends on $\mu$, as
shown on the table. Figure $1(a,b,c,d)$ show  the  distributions  of  ${\bf
\tilde{p}_\perp \cdot \Re\tilde{f}(\tilde{p}_\perp )}$ for typical cases of
$\{k,p_\parallel ,\kappa\}$ values. The Figs. 1(a,b) are for bremsstrahlung
and  Figs.  1(c,d)  are  for ${\bf aws}$ with the parameter values shown in
figures. The black curves show the results from iterations for  zero  baryo
chemical  potential  and the blue curves show for the case of $\mu=1.0$GeV.
As shown in these figures, the $A_i$ values  do  not  change  significantly
with  $\mu$.  The  $A_i$  values  are  generally  less  than unity for most
parameter values for ${\bf aws}$ case. The empiricism has been derived from
a study of several plots of the amplitude function for  various  parameters
and   at   various   stages   during   iterations.   The  resulting  $2{\bf
\tilde{p}_\perp}\cdot\Re   {\bf   \tilde{f}}({\bf    \tilde{p}_\perp    })$
distribution  from  the  iteration method can be used to perform the photon
rate calculation using Eq.(4), though the method is not efficient.  Results
of  this  method  for  low  photon  momentum and high $p_\parallel $ values
become unreliable when the convergence is poor, though the convergence  can
be  improved  by  standard  methods.  However, higher $p_\parallel $ values
($p_\parallel \ge 10T$) contribute lesser to the emission rate owing to the
population functions. Therefore, in the  following  we  adopt  a  different
method for the photon emission rates.

\par
\noindent
We  have  followed  the  variational  approach  discussed in detail in Eqs.
[4.20-4.29] of \cite{arnold2} for solving the integral equation.  We  found
that   this   method   is   remarkably  simple  to  solve  for  the  $2{\bf
\tilde{p}_\perp}\cdot\Re   {\bf   \tilde{f}}({\bf    \tilde{p}_\perp    })$
distributions,  though  highly  intensive  in computing time. The choice of
trial functions  of  \cite{arnold2}  do  not  allow  the  integrals  to  be
analytically tractable and hence can be computed only numerically. Further,
proper  choice  of  the  scale  constant of the variational method $A_v$ is
necessary for use in these trial  functions.  The  $A_v$  value  should  be
around  the  peak  position  of  the  $2{\bf  \tilde{p}_\perp}\cdot\Re {\bf
\tilde{f}}({\bf \tilde{p}_\perp })$  distributions.  For  this  purpose,  a
constraint integral has been given in Eq. {4.24} of \cite{arnold2} . In the
present work, we simply this approach as follows.\\

(i)  The scale constants ($A_v$) are taken from empirical expression of Eq.
8,  ${\it  i.e.,}~  A_v(p_\parallel   ,k)=A_i^2$   for   brensstrahlung   ,
~$A_v(p_\parallel ,k)=A_i$ for $\bf{aws}$.

(ii)  These $A_i$ values are already shown to be weakly dependent on baryon
density. Therefore the scale constants are taken same as for the  $\mu  =0$
case.

(iii)  The  collision kernel of Eq. (6) has been replaced by the analytical
forms as mentioned in the last section.

\par
\noindent
These  three  considerations grossly simplify the problem and bring out the
full advantage of the variational method suggested  in  \cite{arnold2}.  We
have    calculated    the   $2{\bf   \tilde{p}_\perp}\cdot\Re\tilde{f}({\bf
\tilde{p}_\perp})$     distributions     for      various      sets      of
\{$p_\parallel,k,\kappa$\}.  We  compared  these  with the iteration method
mentioned earlier. These distributions are shown by red colored  curves  in
Figs.  1(a-d)  using  the  dimension  ($N_r)$~  12,  of  the  set  of trial
functions. The pink curve is the corresponding  distributions  for  $\mu=1$
GeV  case. As shown in the figures, the agreement between these two methods
is good. This agreement remains valid even  for  low  value  of  $N_r$.  At
finite  baryon  density,  use  of  dimension  $N_r  =  5  ~or~  6 $ is just
sufficient for the photon emission rate calculations using $A_v(p_\parallel
,k)$. Further, for a choice of $N_r$=12 or 15, these distributions will not
be sensitive to the $A_v$ values  used.  This  is  expected  in  any  basis
expansion  method  for  sufficiently  large  dimension  of the model space.
Therefore,  prudent  choice  of  set  of  functions  and  parameters  seems
necessary  to get converging results only for lesser dimensions that render
computations easy. This has been verified by  comparing  the  distributions
for  $N_r$=4  using  $A_v(p_\parallel ,k)$ for a few $\{p_\parallel ,k\}$,
with the results for $N_r$=12 or 15 using any fixed value of $A_v$. It  has
been  observed  that  use  of any constant value of $A_v$ with $N_r$=8 also
gives acceptable results, provided the constant $A_v$ value is neither  too
large  nor  small  as compared to the peak positions. It can be larger by a
factor two and can be smaller by a factor of half than the value  given  by
the  prescription  for $A_v(p_\parallel ,k)$. A priori, the $A_v$ value for
use in calculations for any arbitrary case is not known and  therefore  the
aforesaid prescription serves as a general guideline.

\vspace {1.cm}

\par
\noindent
The  photon emission rates have been calculated for bremsstrahlung and $\bf
aws$ using the variational approach  for  various  values  of  $\mu$  ,  as
denoted  by  ${\cal  R}_{b},{\cal  R}_{a}$ in Eq. 4. The emission rates for
effective two loop processes without LPM effects have also been  calculated
using  Eq.  1  and  are  denoted by ${\cal{R}}^{0}_{b},{\cal{R}}^0_{a}$. We
obtained the suppression factors as defined by the ratio of the rates  from
these  two  methods.  However  for  each  $\mu$  value,  we  normalised the
suppression factors to unity for the $\bf aws$  process  at  very  low  $k$
value  (  $k  \approx 0.2T$). The bremsstrahlung suppression versus $k$ for
all $\mu$ values are asymptotically normalised to zero density case,  ${\it
i.e.,}$  at  ~$k  \sim  20T$. These normalization factors are obtained from
$f_b(\mu),f_a(\mu)$ defined in the following equations. The normalised  (to
zero  density) suppression factors ($S_b,S_a$) as functions of $k,\mu $ are
thus defined as,

\begin{equation}
S_b(k,\mu )= \left(\frac{{\cal{R}}_b}{{\cal{R}}^0_b}\right)_k \frac{f_b(\mu =0)}{f_b(\mu)}
~~~~\mbox{with}~~~~f_b(\mu)= \left(\frac{{\cal{R}}_b}{{\cal{R}}^0_b}\right)_{k=20T}
\end{equation}

\begin{equation}
S_a(k,\mu )= \left(\frac{{\cal{R}}_a}{{\cal{R}}^0_a}\right)_k \frac{f_a(\mu =0)}{f_a(\mu)}
~~~~\mbox{with}~~~~f_a= \left(\frac{{\cal{R}}_a}{{\cal{R}}^0_a}\right)_{k=0.2T}
\end{equation}

\noindent
These  normalised  suppression  factors are shown in Figs. 2(a,b), for each
$\mu$ (quark chemical potential) as captioned in the figures.  The  effects
of  baryon density on these suppression factors is weak only for low baryon
density and the qualitatively these are very similar to the  results  shown
in  \cite{arnold2}.  Our  suppression  factors  agree  quite  well with the
results of \cite{arnold2} for baryon free case. The  normalization  factors
$f_b,f_a$ are shown in the table as a function of quark chemical potential.
The  values of various thermal masses and $\kappa$ values are also shown in
the table. As seen in the table, the normalization factors $f_a$ for  ${\bf
aws}$  are very close to 1.0 showing that the baryon density has no effect.
The ${\bf aws}$ suppression factors  in  Fig.  2(b)  also  show  only  weak
dependence on baryon density. Therefore in total the ${\bf aws}$ process is
insensitive  to  baryon density and one can use the suppression factors for
zero density case for use in photon spectrum calculations. However, for the
bremsstrahlung case the density effect is not totally weak. It can be  seen
from  the  table  that  the  $f_b(\mu)$ value varies considerably for large
chemical potential values. The total suppression curves as  a  function  of
$k$  for the bremsstrahlung and ${\bf aws}$ at a given $\mu$ are thus given
by multiplying with  $\frac{f_{b,a}(\mu)}{f_{b,a}(\mu=0)}$  the  respective
curves  in  Fig. 2(a,b) to undo the normalization. For low density, one can
use the suppression factors as a function of  $k$  for  zero  density  case
multiplied        by        a        density        dependent       factors
$\frac{f_{b,a}(\mu)}{f_{b,a}(\mu=0)}$.  It  may be of interest to know from
the table that $\kappa f_{b}(\mu)$  is  roughly  a  constant  (=0.214)  and
therefore  the density dependent factor can be chosen as $\kappa_0/\kappa$.
These calculations can be extended to  an  important  case  of  the  plasma
conditions  at  RHIC  collision,  where  the  $\kappa$  values  can be very
different from the values in present study.

\vspace {1.cm}
\par
\par
\noindent
{\bf Conclusion}

\noindent
The  photon  emission  rates  from the quark gluon plasma have been studied
considering LPM effects. Self-consistent iterations method has been used to
solve   the   integral   equation   for   the   ${\bf    \Re\tilde{f}}({\bf
\tilde{p}_\perp})$ distributions. The peak positions of these distributions
have  been  fitted  by  an empirical expression for various parameter sets.
These peak positions are observed to be rather insensitive  to  the  baryon
density.   The   variational   approach   has   been   adopted   with  some
simplifications. Analytical forms for collision kernel have  been  used  in
both  the iterations and variational methods. The scale constants have been
used from empirical expressions  from  iteration  method.  The  suppression
factors  for  bremsstrahlung and $\bf aws$ have been obtained as a function
of photon energy at different baryon densities. The normalised  suppression
factors  are  qualitatively  similar  to  the zero baryon density case. The
normalization factors show the density dependence for  bremsstrahlung.  The
${\bf aws}$ process did not show significant density dependence.

\vspace {1.cm}

\par

\par
\noindent
{\bf Acknowledgements}

\noindent
We  acknowledge  the  fruitful  discussions with Dr. S. Kailas and also for
initiating us to the use of parallel processors computing system  for  this
study.  Dr. A.K. Mohanty is acknowledged for initiating me to these studies
and  his  immense  contribution  in  my  career.  Dr.  A.K.   Mohanty   and
collaborators  are  specially  thanked.  We thank Dr. A. Navin for his keen
interest in this work. We thank  especially  the  head  computer  division,
BARC.  Our  sincere thanks to the staff of computer division P.S. Dekne, K.
Rajesh, Jagdeesh, P.  Saxena  and  also  the  operating  staff  for  making
available the BARC Anupam parallel processors computing system and constant
co-operation  during  this  study. The iteration part of the present study,
the  resulting  peak  position  parameterization,  the  emission   spectrum
calculations  have  been  feasible,  thanks  to  this  parallel  processors
facility.

%\end{document}
%\newpage
%%%%%%%%%%%%%%%%%%space-time expansion%%%%%%%%%%%%%%%%%%%%%%%%%
%\newpage
\begin{figure}[!ht]
%\vspace{-1.0cm}
%\hspace{-1.cm}
\begin{center}{
\hspace{-.5cm}
\begin{minipage}{22.cm}
\psfig{figure=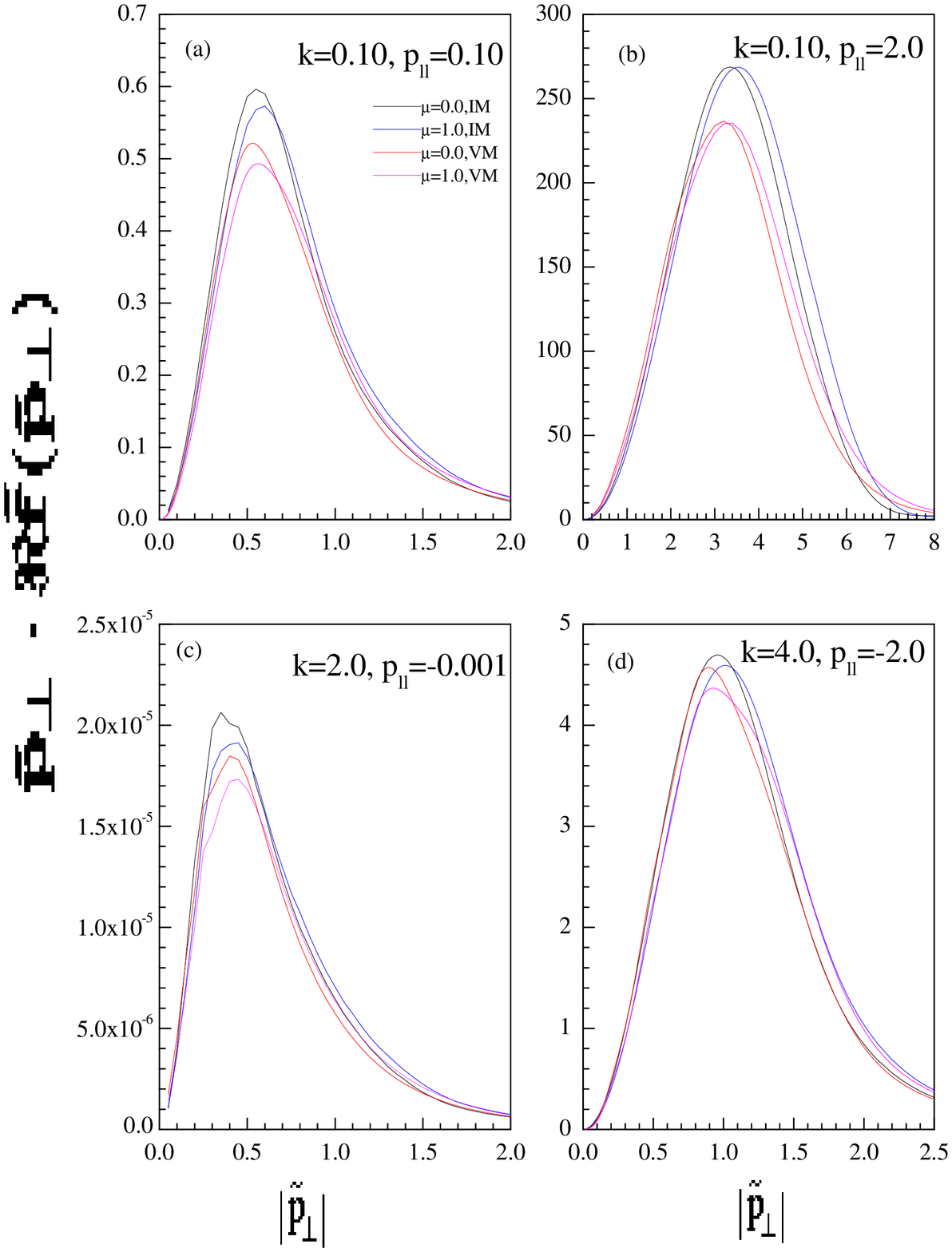,height=22.cm,width=16cm}
%\vspace{-.50cm}
\end{minipage}
\caption{   The   $2{\bf   \tilde{p}_\perp}\cdot\Re   {\bf  \tilde{f}}({\bf
\tilde{p}_\perp  })$  distributions  from   iteration   method   (IM)   for
bremsstrahlung  (Figs.(a,b))  and  ${\bf aws }$ (Figs.(c,d)) processes. The
black curves are for the case of zero density and the blue curves  are  for
finite  density  with  1.0  GeV quark chemical potential. The corresponding
cases from variational method (VM) are shown in red  and  pink  color.  The
${p_\parallel},k$ values (in GeV) used are shown in figures.}
}\end{center}
%\vspace {-0.25cm}
\end{figure}
%\newpage
\begin{figure}[!ht]
%\vspace{-1.0cm}
%\hspace{-1.cm}
\begin{center}{
\hspace{-2.5cm}
\begin{minipage}{17.cm}
\psfig{figure=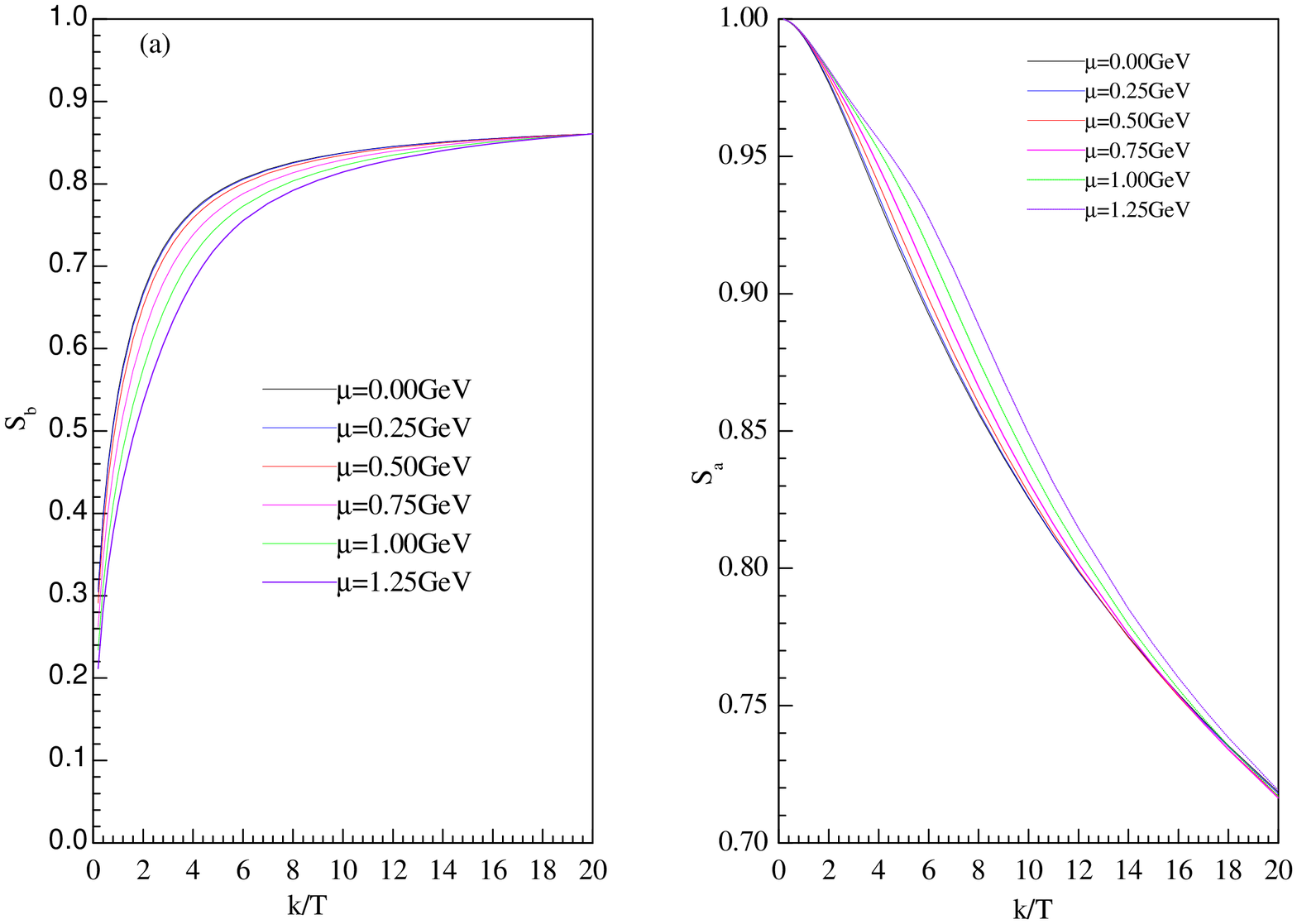,height=12.cm,width=19.cm}
%\vspace{-.50cm}
\end{minipage}
\caption{(a)  bremsstrahlung and (b) ${\bf aws}$ suppression factors versus
photon energy. The suppression factors have been normalised to zero density
case  asymptotically  for  bremsstrahlung   and   for   ${\bf   aws}$   the
normalization  is  at  $k=0.2T$.  The true curves at nonzero density can be
obtained from the normalization factors given in  table,  as  discussed  in
text.}
}\end{center}
%\vspace {-0.25cm}
\end{figure}
\begin {center}{
\begin{table}
\begin{tabular}
{|c|c|c|c|c|c|c|c}
\hline
 $\mu$ & $m_g^2/m_\infty^2$  & $\kappa$ & $m_\infty^2$ & $m_d^2$ & $f_b$ & $f_a$ \\
\hline
  0.00 & 1.33333 &  0.25000 &  0.05236 &  0.20944 &  0.85530    & 0.99382  \\
  0.10 & 1.32802 &  0.25100 &  0.05321 &  0.21199 &  0.85250    & 0.99349  \\
  0.25 & 1.30267 &  0.25589 &  0.05767 &  0.22535 &  0.83952    & 0.99192  \\
  0.50 & 1.23720 &  0.26943 &  0.07358 &  0.27310 &  0.80634    & 0.98780  \\
  0.75 & 1.17435 &  0.28385 &  0.10011 &  0.35268 &  0.77023    & 0.98379  \\
  1.00 & 1.12717 &  0.29572 &  0.13724 &  0.46408 &  0.73468    & 0.98074  \\
  1.25 & 1.09434 &  0.30460 &  0.18500 &  0.60736 &  0.70158    & 0.97860  \\
  1.50 & 1.07173 &  0.31102 &  0.24332 &  0.78233 &  0.67098    & 0.97711  \\
  1.75 & 1.05588 &  0.31569 &  0.31236 &  0.98944 &  0.64143    & 0.97607  \\
  2.00 & 1.04454 &  0.31912 &  0.39182 &  1.22782 &  0.61527    & 0.97532  \\
\hline
\end {tabular}
\end{table}
}\end {center}
\end{document}